\documentclass[superscriptaddress, preprint,amsmath,amssymb,prb]{revtex4-1}
\usepackage{graphicx}
\usepackage{dcolumn}
\usepackage{multirow}
\usepackage{booktabs}
\usepackage{bm,color}
\usepackage{braket}

\begin{document}

\title{Spin Hall effect emerging from a noncollinear magnetic lattice without spin-orbit coupling}

\author{Yang Zhang}
\affiliation{Max Planck Institute for Chemical Physics of Solids, 01187 Dresden, Germany}
\affiliation{Leibniz Institute for Solid State and Materials Research, IFW Dresden, 01069 Dresden, Germany}

\author{Jakub \v{Z}elezn{\'y}}
\affiliation{Max Planck Institute for Chemical Physics of Solids, 01187 Dresden, Germany}

\author{Yan Sun}
\affiliation{Max Planck Institute for Chemical Physics of Solids, 01187 Dresden, Germany}

\author{Jeroen van den Brink}
\affiliation{Leibniz Institute for Solid State and Materials Research, IFW Dresden, 01069 Dresden, Germany}
\affiliation{Institute for Theoretical Physics, TU Dresden, 01069 Dresden, Germany}


\author{Binghai Yan}
\email{binghai.yan@weizmann.ac.il}
\affiliation{Department of Condensed Matter Physics, Weizmann Institute of Science, 7610001 Rehovot, Israel}

\begin{abstract}
The spin Hall effect (SHE), which converts a charge current into a transverse spin current,
has long been believed to be a phenomenon induced by the spin--orbit coupling.
Here, we identify an alternative mechanism to realize the intrinsic SHE through a noncollinear magnetic structure that breaks the spin rotation symmetry.
No spin--orbit coupling is needed even when the scalar spin chirality vanishes, different from the case of the topological Hall effect and topological SHE reported previously.
In known noncollinear antiferromagnetic compounds Mn$_3X$ ($X=$ Ga, Ge, and Sn), for example,
we indeed obtain large spin Hall conductivities based on \textit{ab initio} calculations.
\end{abstract}


\maketitle

\section{introduction}\label{intro}

The spin Hall effect (SHE)~\cite{Sinova2015} is one of the most important ways to create and detect spin currents in the field of spintronics, which aims to realize low-power-consumption and high-speed devices.~\cite{Wolf2001}
It converts electric currents into transverse spin currents and vice versa.
The SHE in materials is generally believed to rely on spin--orbit coupling (SOC)~\cite{Sinova2015,Maekawa2012,Hoffmann2013,Gradhand2012}.
In typical SHE devices, the generated spin current is injected into a ferromagnet (FM) and consequently switches its magnetization via the spin-transfer torque~\cite{Miron2011,Liu2012} or drives an efficient motion of magnetic domain walls.~\cite{Parkin2015,Yang2015}

The SHE is conceptually similar to the well established anomalous Hall effect
(AHE). In recent decades, the understanding of the intrinsic
AHE~\cite{Nagaosa2010} and intrinsic SHE~\cite{Murakami2003,Sinova2004} was
significantly advanced based on the concept of the Berry phase,~\cite{Xiao2010}
which originates directly from the electronic band structure. Although the AHE requires the existence of SOC in a FM, it also appears in a non-coplanar magnetic lattice  without SOC, where an electron acquires a Berry phase by hopping through sites with a noncoplanar magnetic structure (nonzero scalar spin chirality)~\cite{Ohgushi2000,Taguchi2001}, later referred as the topological Hall effect (THE)~\cite{Bruno2004}.
Thus, in experiment the THE-induced Hall signal is
considered as a signature of the skyrmion phase with chiral spin texture~\cite{Neubauer2009,Kanazawa2011}.
Provoked by the THE, recent numerical simulations of the spin scattering by a
single skyrmion indicated the presence of a finite SHE even without SOC~\cite{Yin2015,Buhl2017,ndiaye2017topological}, which is termed as a topological SHE.
Thus, the topological SHE has been presumed to stem from the Berry phase due to the nonzero spin chirality of the skyrmion.
However, the origin of the spin current is illusive in the topological SHE,
for it is hard to separate it from the spin-polarized charge current of the THE.
Here, we pose new questions one step further.
Is the skyrmion-like spin texture (nonzero scalar spin chirality) always necessary to generate a SHE without SOC? What is the generic condition for a SHE without SOC?

In this article, we propose a mechanism to realize the SHE with the noncollinear magnetic structure but without SOC. The crucial role of SOC is to break the spin rotational symmetry (SRS) in SHE. Alternatively, it is known that common noncollinear magnetic textures can also violate the SRS, thus resulting in the SHE. Different from the THE and topological SHE in symmetry, such an SHE appears universally for the noncollinear magnetic lattice, regardless of the scalar spin chirality. For example, it can even emerge in a coplanar magnetic structure where the scalar spin chirality is zero.
 Here, we first prove the principle from the symmetry analysis in a simple lattice model. Then, we demonstrate the existence of a strong SHE in several known materials Mn$_3X$ ($X=$Ga, Ge, and Sn)\cite{Nayak2016,Nakatsuji2016} by \textit{ab initio} calculations without including SOC.



\begin{figure}[htbp]
\begin{center}
\includegraphics[width=0.8\textwidth]{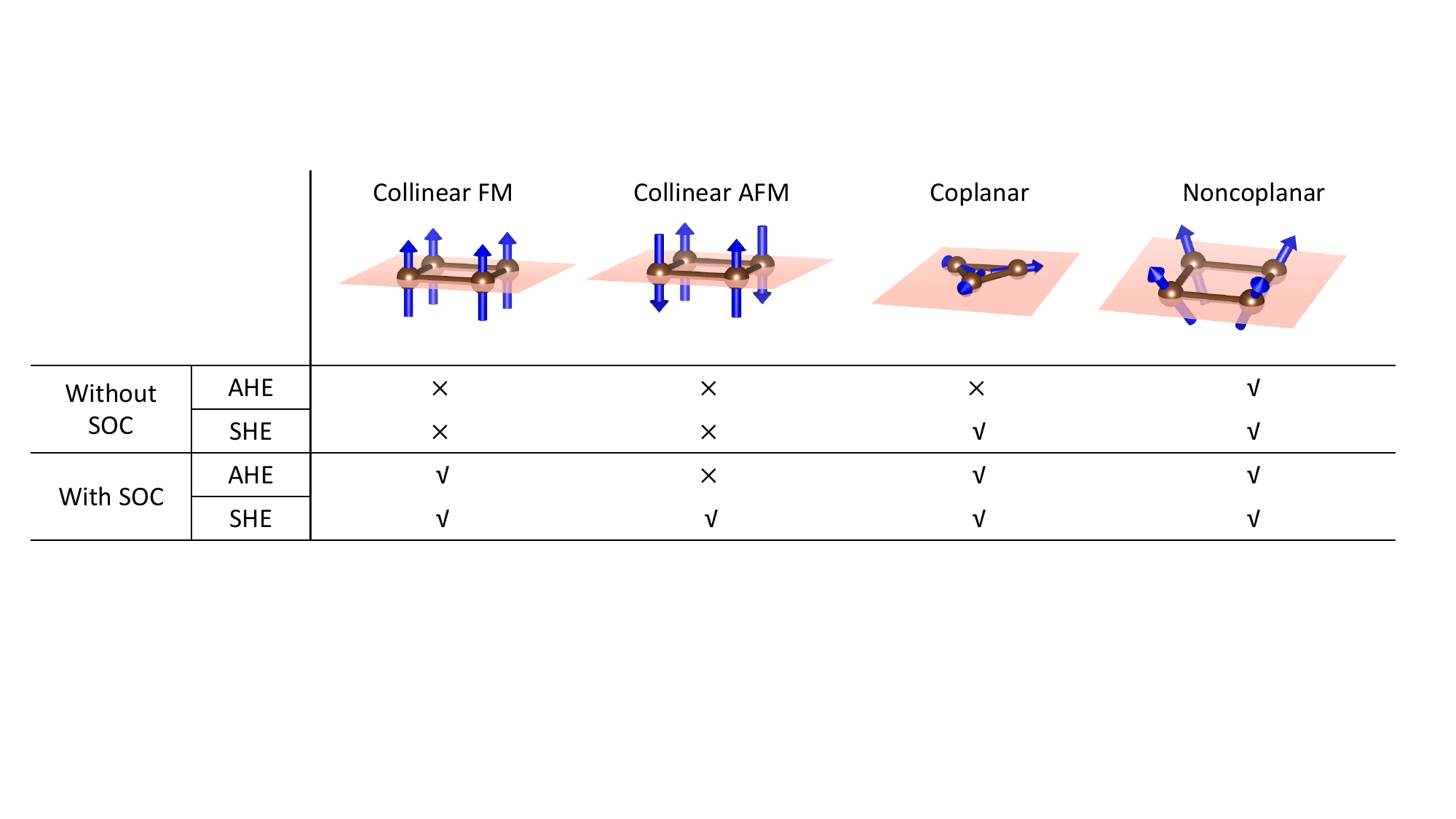}
\end{center}
\caption{
Symmetry conditions for the existence ($\surd$) or absence ($\times$) of AHE and SHE
 in collinear FM, collinear AFM, coplanar, and noncoplanar magnetic lattices with and without SOC.
Note that the SHE is symmetry allowed when the magnetic ordering is coplanar
  (but noncollinear) even without SOC (see text).
}
\label{AHESHE}
\end{figure}

\section{results}
\textit{Double-exchange model and symmetry analysis --}
The existence of the SHE and AHE in metals is determined by symmetry (in
insulators apart from symmetry also the topology of the electronic structure is
important). The symmetry of magnetic systems is normally described in terms of
magnetic space groups, which contain, apart from the spatial symmetry
operations, also the time-reversal symmetry operation. In absence of SOC (or
other terms in the Hamiltonian that couple the magnetic moments to the lattice such as the shape anisotropy), however, the symmetry of the magnetic systems is higher than that contained in the magnetic space groups since the spins can be rotated independently of the lattice. This can be illustrated by considering  the following minimal Hamiltonian
\begin{equation}
\label{sd-model}
\begin{aligned}
  H = t\sum_{<ij>\alpha}c^{\dag}_{i\alpha}c_{j_\alpha}  -
  J\sum_{i \alpha \beta} (\bm{\sigma} \cdot \bm{n}_i)_{\alpha \beta} c^{\dag}_{i\alpha}c_{i\beta},
\end{aligned}
\end{equation}
known as the double-exchange model ($s$-$d$ model)~\cite{Zener1951,Anderson1955,Gennes1960} which describes itinerant $s$ electrons interacting with local $d$ magnetic moments. We assume that magnetic moments are only contributed by the spin degrees of freedom. Here, $\alpha$ and $\beta$ stand for spin up and spin down, respectively. The
first term is the nearest neighbor hopping term with $<ij>$ denoting the nearest
neighbor lattice sites. In the second term, $J$ is the Hund's coupling strength
between the conduction electron and the on-site spin moment, $\bm{\sigma}$ is the vector of Pauli matrices, and $\bm{n}_i$ is the spin magnetic moment on site $i$.
The magnetic texture is defined by the pattern of $\bm{n}_i$.
In such a Hamiltonian, spin rotation only rotates the magnetic moments $\mathbf{n}_i$.
The corresponding symmetry groups are thus formed by combining the magnetic space groups with spin rotations.\cite{Litvin1974,Brinkman1966} Such symmetry groups are referred to as spin-space groups. They apply generally for non-interacting Hamiltonians in absence of spin-orbit coupling.

We focus here only on the intrinsic contribution to the AHE and SHE, however, the other (extrinsic) contributions have the same symmetry and thus the symmetry discussion in the following is general. The intrinsic AHE and intrinsic SHE are well characterized via the Berry curvature formalism.~\cite{Xiao2010, Nagaosa2010, Sinova2015, Gradhand2012}
The anomalous Hall conductivity(AHC) $\sigma_{\alpha \beta}$ can be evaluated by
the integral of the Berry curvature $\Omega^n_{\alpha\beta}(\mathbf{k})$ over the Brillouin zone (BZ)
for all the occupied bands, where $n$ is the band index.
It should be noted that this method can also be applied to the THE, although it is commonly interpreted using the real space spin texture.
Here, $\sigma_{\alpha \beta}$ corresponds to a $3\times3$ matrix and
indicates a transverse Hall current $j_\alpha$ generated by a longitudinal electric field $\mathbf{E}$, which satisfies
$J_\alpha=\sigma_{\alpha \beta}E_\beta$. Within a linear response, Berry curvature can be expressed as
\begin{equation}
\label{AHC}
\begin{aligned}
\Omega_{n,\alpha\beta}(\vec{k})= 2i\hbar^2 \sum_{m \ne n} \dfrac{\bra{\psi_{n\mathbf{k}}}\hat
  v_{\alpha}\ket{\psi_{m\mathbf{k}}}\bra{\psi_{m\mathbf{k}}}\hat v_{\beta}\ket{\psi_{n\mathbf{k}}}}{(E_{n}(\vec{k})-E_{m}(\vec{k}))^2},
\end{aligned}
\end{equation}
where $n$ and $m$ are band indices, and $\psi_{n\mathbf{k}}$ and $E_{n\mathbf{k}}$ denote the Bloch wave functions and eigenvalues, respectively, and $\hat{\mathbf{v}}$ is the velocity operator. Replacing the velocity operator with the spin current operator
$\hat J_{\alpha}^{\gamma}=\dfrac{1}{2} \{ \hat v_{\alpha},\hat s_{\gamma}\}$,
where $\hat s_{\gamma}$ is the spin operator, we obtain the spin Berry
curvature and corresponding spin Hall conductivity (SHC),
\begin{equation}
\label{SHC}
\begin{aligned}
\sigma_{\alpha \beta}^{\gamma}&= \dfrac{e}{\hbar}\sum_n
\int_{BZ}\dfrac{d^3\mathbf{k}}{(2\pi)^3}f_n(\mathbf{k})\Omega_{n,{\alpha \beta}}^{\gamma}(\mathbf{k}), \\
\Omega_{n,\alpha \beta}^{\gamma}(\mathbf{k})&= 2i \hbar^2 \sum_{m \ne n} \dfrac{\bra{\psi_{n\mathbf{k}}} \hat
J_{\alpha}^{\gamma}\ket{\psi_{m\mathbf{k}}}\bra{\psi_{m\mathbf{k}}}\hat{v}_{\beta}\ket{\psi_{n\mathbf{k}}}}{(E_{n\mathbf{k}}-E_{m\mathbf{k}})^2}.
\end{aligned}
\end{equation}
The SHC ($\sigma_{\alpha \beta}^{\gamma}$; $\alpha,\beta,\gamma=x,y,z$)
is a third-order tensor ($3\times3\times3$) and represents the spin current ${J}_{s,\alpha}^{\gamma}$
generated by an electric field $\mathbf{E}$ via
${J}_{s,\alpha}^{\gamma}=\sigma_{\alpha \beta}^{\gamma}{E}_{\beta}$,
where ${J}_{s,\alpha}^{\gamma}$ is a spin current flowing along the $\alpha$-direction with the
spin-polarization along the $\gamma$-direction, and $f_n(\mathbf{k})$ is the
temperature dependent Fermi-Diract distribution.

We know that AHE vanishes while SHE remains if the time-reversal symmetry (operator $T$) exists in the system. In Eq.~\ref{AHC}, $T$ reverses the velocities $\hat v_{\alpha,\beta}$ and brings an additional ``$-$'' sign by the complex conjugate. Thus, $\sigma_{\alpha\beta}=0$  owing to $\Omega_{n,\alpha\beta}(\vec{k})=-\Omega_{n,\alpha\beta}(\vec{-k})$. In contrast, In Eq.~\ref{SHC}, $T$ generates one more ``$-$'' sign by reversing the spin in ${J}_{\alpha}^{\gamma}$. Then, $\sigma_{\alpha \beta}^{\gamma}$ can be nonzero since $\Omega_{n,\alpha\beta}^{\gamma}(\vec{k})$ is even in $k$-space.
In a magnetic system without SOC, $T$ is broken, but a combination of $T$ and a spin rotation (operator $S$) can still be a symmetry. For example, a coplanar magnetic system shows a $TS$ symmetry, in which $S$ rotates all spins by $180^\circ$ around the axis perpendicular to the plane. Since $S$ does not act on $\hat v_{\alpha,\beta}$, $TS$ causes vanishing $\sigma_{\alpha\beta}$ just as $T$ alone. In a general noncoplanar magnetic lattice, the $TS$ symmetry is naturally broken, because one cannot find a common axis about which all spins can be rotated $180^\circ$ at the same time, and thus the AHE can exist without SOC.

The situation is different for the SHE since ${J}_{\alpha}^{\gamma}$ in Eq.~\ref{SHC} contains an additional spin operator. As a consequence, (assuming that $S$ is a rotation around the $z$ axis) $TS$ forces $\Omega_{n, \alpha \beta}^{x/y}(\vec{k})$ to be odd where spin $\hat s_{x}$ or $\hat s_y$ is reversed by $TS$, but $\Omega_{n, \alpha\beta}^{z}(\vec{k})$ to be even where $\hat s_{z}$ is unchanged by $TS$. Then, one can obtain zero $\sigma_{\alpha \beta}^{x/y}$ but nonzero $\sigma_{\alpha \beta}^{z}$. In a collinear magnetic lattice there exists more than one spin rotation $S$ such that $TS$ is a symmetry of the system and thus all of the $\sigma_{\alpha \beta}^{\gamma}$ components have to vanish. Therefore, we can argue that SHE can exist without SOC in general noncollinear magnetic lattices, regardless of FM, AFM, or the scalar spin chirality (coplanar or noncoplanar). In contrast, the AHC is zero for a coplanar magnetic lattice (zero scalar spin chirality), since $TS$ acts as $T$ alone in Eq.~\ref{AHC}.

When SOC is included, SHE is allowed by symmetry in any crystal,\cite{Seemann2015} while the AHE on the other hand can be present in magnetic systems that are not symmetric under time reversal combined with a translation or inversion (for example, a conventional collinear AFM). We summarize the necessary conditions for the existence of AHE and SHE in systems with and without SOC in Fig. \ref{AHESHE}.

\begin{figure}[htbp]
\begin{center}
\includegraphics[width=0.8\textwidth]{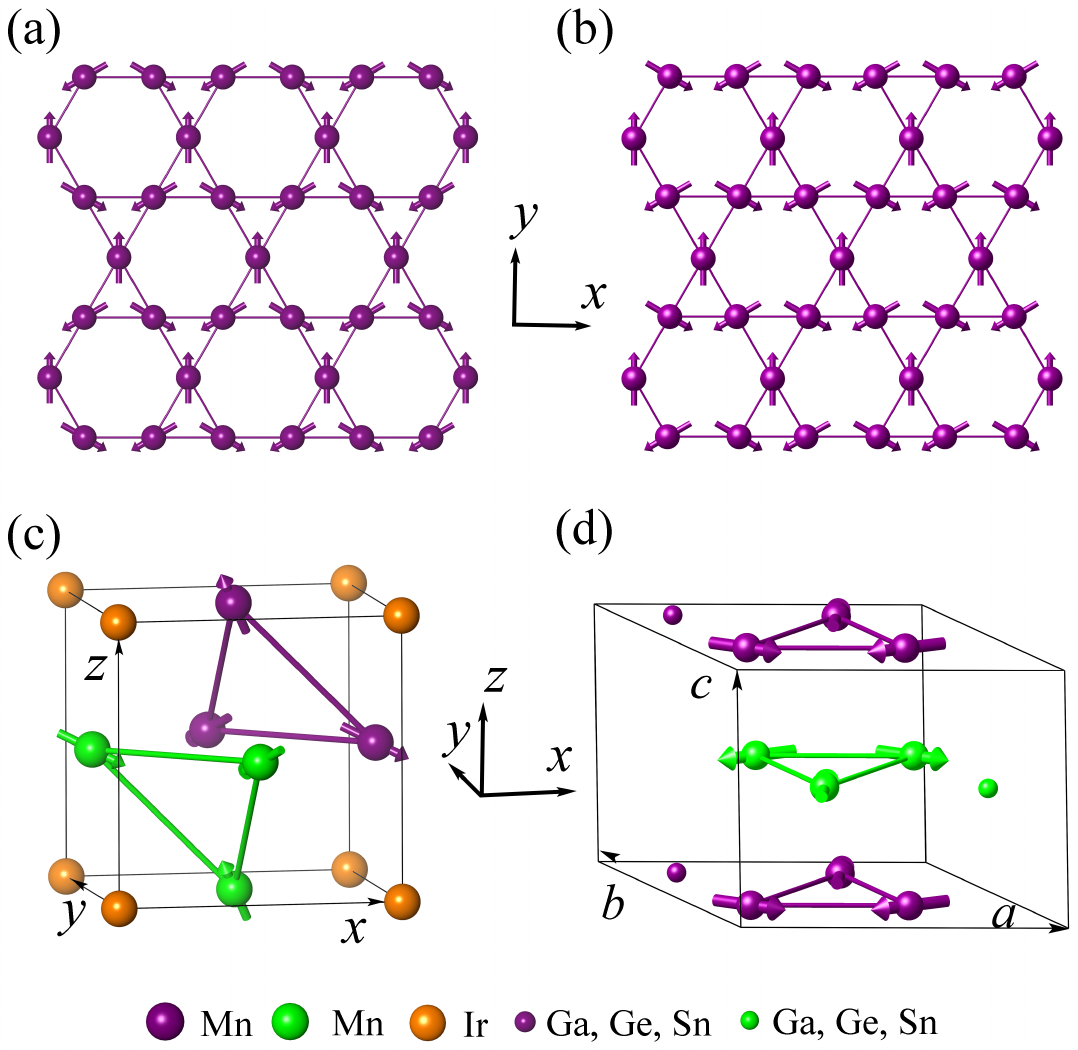}
\end{center}
\caption{Noncollinear order in kagome lattice and the magnetic structure of Mn$_3$Ir and Mn$_3$X (X = Ga, Ge, and Sn).
(a) The $q=0$ order in the kagome lattice, with magnetic moments located as 2D Mn plane in Mn$_3$Ir,
(b) Two-fold spin rotation around $y$-axis of configuration (a), corresponding to Mn planes in Mn$_3$X (X = Ga, Ge, Sn),
(c) The face-centered cubic crystal structure of Mn$_3Ir$,
(d) The hexagonal crystal structure of Mn$_3$X (X = Ga, Ge, and Sn).
}
\label{stru}
\end{figure}

To demonstrate that the SHE can indeed be nonzero without SOC, we consider the $s$--$d$ Hamiltonian \eqref{sd-model} projected on a kagome lattice with the so-called $q=0$ magnetic order, shown in Fig. \ref{stru}(a). Such a coplanar AFM order is well studied in theory~\cite{Ohgushi2000,Chen2014Mn3Ir} and appears in many realistic materials even at room temperature, for example Mn$_3X$ ($X=$ Ir, Ga, Ge, and Sn)~\cite{zhang2013,Kubler2014,Nakatsuji2016,Nayak2016,Zhang2016Mn3Ir,Zhang2017} as we discuss in the following. For comparison, the SOC effect is also considered by adding to $H$ in Eq.~\ref{sd-model},
\begin{align}
 H_\text{so} = it_2\sum_{<ij>\alpha\beta}v_{ij}(\bm{\sigma} \cdot
  \bm{n}_{ij})_{\alpha\beta}c^{\dag}_{i\alpha}c_{j\beta},
\end{align}
where $v_{ij}$ is the antisymmetric Levi-Civita symbol and $n_{ij}$
  are a set of coplanar vectors anticlockwise perpendicular to the lattice
vector $R_{ij}$, as defined in Ref. \onlinecite{Chen2014Mn3Ir} and $t_2$ is the SOC strength.

We first analyze the symmetry of the SHC tensor for the $q=0$ magnetic order.
Note that we use the Cartesian coordinate systems defined in Fig. \ref{stru}. As discussed above, the existence of the $TS$ symmetry leaves only $\sigma_{\alpha \beta}^{z}$ terms in the absence of SOC. Further, the combined symmetry $T M_x$, in which $M_x$ is the mirror reflection along $x$ and flips $\hat s_z$ and $\hat v_x$ in Eq.~\ref{SHC}, leads to $\sigma_{xx}^{z}=\sigma_{yy}^{z}=0$. We further obtain only two nonzero SHC tensor element $\sigma^z_{xy}=-\sigma^z_{yx}$ by considering the three-fold rotation around $z$.
The magnetic order shown in Fig. \ref{stru}(b) will also be relevant for the discussion in the following. This  magnetic configuration differs from the $q=0$ case only by a two-fold spin rotation around the $y$-axis and thus, without SOC its symmetry is exactly the same as that of the $q=0$ case.

\begin{figure}[htbp]
\begin{center}
\includegraphics[width=0.8\textwidth]{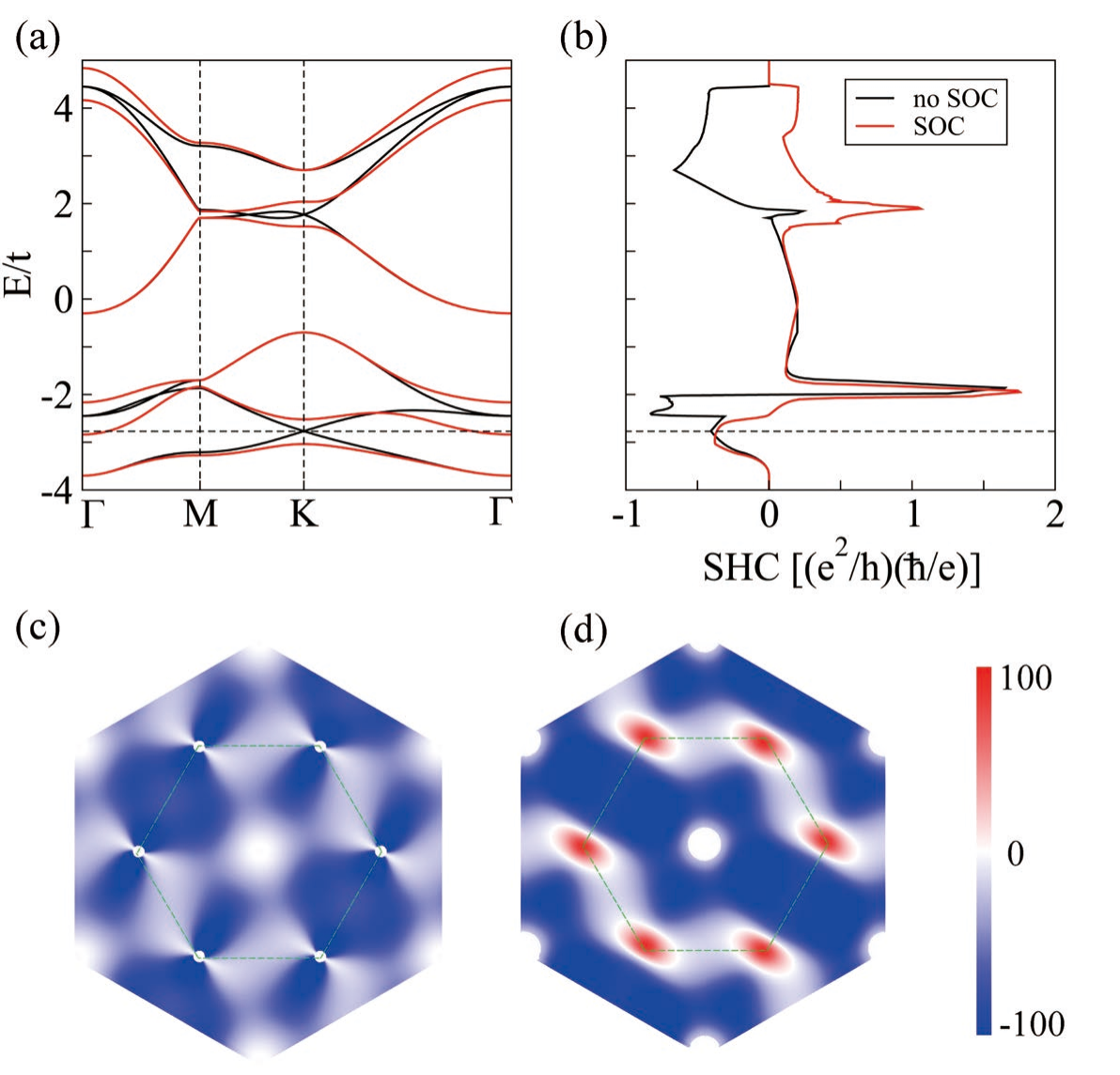}
\end{center}
\caption{Electronic band structures $q=0$ order in kagome lattice,
  (a) without and with SOC. (b) Energy-dependent SHC $\sigma^z_{xy}$. (c) Spin
  Berry curvature $\Omega^z_{xy}$ distribution of first BZ at Fermi energy
  $-2.7$ eV (horizontal line) without SOC and (d) with SOC.}
\label{sberry}
\end{figure}

Setting the Hund coupling constant $J=1.7t$ and SOC strength $t_2 = 0$, we
calculate the spin Berry curvature via Eq. ~\ref{SHC}. As expected, we find nonzero SHC $\sigma^z_{xy}$ fully in agreement with the symmetry considerations. Figures \ref{sberry}a and \ref{sberry}b show the band structures with $t_2 = 0$ and $t_2=0.2t$, respectively. One can see that SOC modifies slightly the band structure by gaping some band crossing points such as the BZ corners ($K$). Without SOC, we already observe nonzero $\sigma^z_{xy}$, while adding SOC reduces $\sigma^z_{xy}$ slightly at the Fermi energy that is set between the first and second bands at about --2.7 eV.
We plot corresponding  spin Berry curvature $\Omega^z_{xy}$ in the hexagonal BZ In Figs. \ref{sberry}d and \ref{sberry}e.
Large $\Omega^z_{xy}$ appears in the BZ without SOC, leading to net  $\sigma^z_{xy}$.
The SOC simply brings an extra contribution to $\sigma^z_{xy}$ at the band anti-crossing region around the $K$ point.

\begin{figure}[htbp]
\begin{center}
  \includegraphics[width=0.8\textwidth]{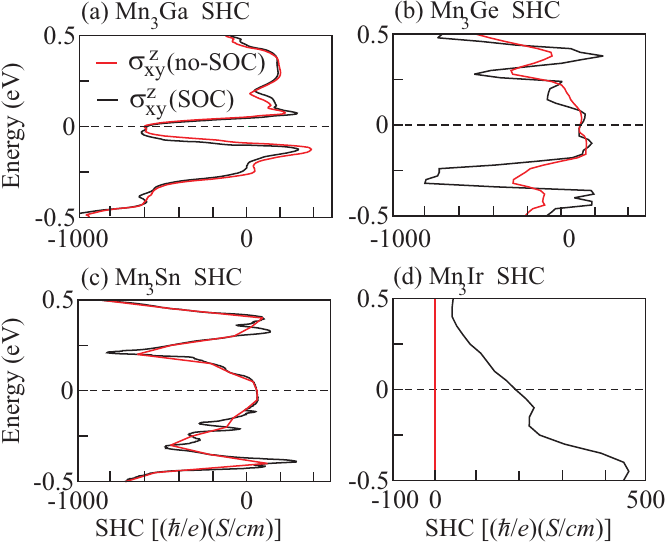}
\end{center}
\caption{
Energy-dependent SHC tensor elements of $\sigma_{xy}^{z}$ with and without SOC for
(a) Mn$_3$Ga, (b) Mn$_3$Ge, (c) Mn$_3$Sn, and (d) Mn$_3$Ir.
The Fermi energy is indicated by the dashed horizontal line.
}
\label{band-Ge}
\end{figure}

\textit{Realistic materials --}
After proving the principle, we now identify materials that show strong SHE with negligible contribution from SOC.
We naturally consider Mn$_3$X (X=Ga, Ge, Sn, and Ir) compounds, since they
exhibit non:w
collinear AFM order at room temperature (the AFM N\'eel temperature is over 365 K). Our recent \textit{ab~ initio} calculations showed a sizable intrinsic SHE by including SOC~\cite{Zhang2017}. Here, we further point out that SHE still presents without SOC and SOC actually plays a negligible effect for SHE in these materials.

The primitive unit cell of Mn$_3$Ga, Mn$_3$Ge and Mn$_3$Sn (space group $P6_3/mmc$, No. 194) includes two Mn$_3X$ planes that are stacked along the $c$-axis according to a ``--AB--AB--'' sequence. Inside each plane, Mn atoms form a kagome-type lattice with Ga, Ge, or Sn lying at the center of the hexagon formed by Mn. Both the \textit{ab initio}
calculation~\cite{zhang2013} and neutron diffraction measurements~\cite{kren1970neutron,Kadar1971, Ohoyama1961} show that the Mn magnetic moments exhibit noncollinear AFM order, where the neighboring moments are aligned at an angle of 120$^\circ$, as in Fig. \ref{stru}(b). Large AHE in room temperature has recently been reported in Mn$_3$Ge and Mn$_3$Sn.~\cite{Nakatsuji2016,Nayak2016} These materials also exhibit other exotic phenomena including the Weyl semimetal phase,~\cite{Yang2016} magneto-optical Kerr effect,~\cite{Feng2015} anomalous Nernst effect,~\cite{Li2016} and topological defects.~\cite{Liu2017}
Distinct from hexagonal Mn$_3X$ compounds, the Mn$_3$Ir (space group $Pm\bar{3}m$, No. 221) crystallizes in a face-centered cubic structure with Mn atoms in the [111] planes forming a kagome lattice with the $q=0$ magnetic order.

The symmetry of the SHE without SOC in these materials can be understood using a similar approach as we used for the 2D kagome lattice. The hexagonal Ga, Ge, and Sn materials can be viewed as stacking versions of the kagome lattice and thus we find that the symmetry of SHE is the same as the 2D kagome lattice, i.e. only $\sigma_{xy}^z = -\sigma_{yx}^z$ is nonzero.
However, we find that SHE must vanish in Mn$_3$Ir without SOC, which is imposed by the higher symmetry of the cubic magnetic lattice. For completeness, we list the tensor matrices without and with SOC for all these compounds in the appendix.


Since the SHC tensor shape imposed by the symmetry has been systematically
investigated for these materials in Ref.~\onlinecite{Zhang2017}, we only discuss
one of the largest SHC tensor elements $\sigma_{xy}^z$ based on the \textit{ab
initio} calculations~\cite{kresse1996} of the SHC. For comparison, we show the
SHC without and together with SOC in Fig.~\ref{band-Ge}. In
the absence of SOC, Ga, Ge, and Sn compounds indeed exhibit nonzero SHC
$\sigma_{xy}^z = -613, 115,$ and $90 ~ (\hbar/e)(\Omega\cdot cm)^{-1}$, respectively, at the Fermi energy.
One can see that SOC induces very few changes in the band structure and thereafter modifies the SHC weakly, especially at the Fermi energy for Ga, Ge, and Sn compounds.
It is intuitive to observe comparable $\sigma_{xy}^z$ values for Ge and Sn compounds, despite the fact that Sn exhibits much larger SOC than Ge. These facts further verifies that the noncollinear magnetic structure, rather than SOC, is dominant for the SHE.
The Ga compound shows an opposite sign in SHC compared to the Ge/Sn compound, since Ga has one valence electron fewer than Ge/Sn and the Fermi energy is lower in Mn$_3$Ga than in Mn$_3$Ge/Mn$_3$Sn.

\section{discussion}

Understanding the role SOC plays in the SHE is important for the fundamental understanding of the SHE, but also for practical reasons. It can help with the search for materials with large SHE since in non-magnetic or collinear magnetic materials, SOC is necessary for SHE and thus the presence of heavy elements is generally required for large SHE.
The SHE without SOC proposed in this work suggests a new strategy to design SHE materials
without necessarily involving heavy elements.
In noncollinear systems, the Rashba
effect can also appear without SOC~\cite{liu2013tunable}. The spin texture in
the band structure may depend sensitively on the real space spin texture.
For example, we found that band structure spin texture is different
between the Kagome lattice and the triangular lattice.

We propose the general, necessary symmetry-breaking requirements (Fig. \ref{AHESHE}) for SHE without SOC. It is worth noting that SHE can become zero without SOC in some noncollinear magnetic lattice where additional symmetries forces the SHE to vanish. For example, we
have shown that in Mn$_3$Ir the SHE vanishes in absence of SOC even though it
has a noncollinear magnetic structure. This is a consequence of its
high-symmetrical cubic structure. Similar situation could happen for AHE without
SOC in a noncoplanar magnetic lattice, such as the AFM skyrmion system~\cite{ndiaye2017topological}.

In conclusion, we have shown that the SHE can be realized by a non-chiral
coplanar magnetic structure without involving the SOC.
The noncollinearity of the magnetic lattice can break the spin rotation symmetry
and consequently allow the existence of SHE. By \textit{ab initio} calculations,
we further predicted that such an SHE without SOC can be observed in noncollinear
AFM compounds Mn$_3X$ ($X = $ Ga, Ge, and Sn).
From our symmetry considerations, an extrinsic SHE
can appear when breaking the spin rotational symmetry. Thus,
we expect the extrinsic effect to also exist in our systems.
Its amplitude will depend on the details of the scattering,
and cannot be estimated without microscopic calculations, though
in general for SHE the intrinsic contribution tends to be the dominant
contribution.
By providing a general theoretical, symmetry based understanding of the SHE,
our work motivates a comprehensive search for SHE materials among noncollinear magnetic systems, that not
necessarily involve heavy elements.
In addition, the close relation between the SHE and the magnetic order suggests
that the SHE may be used vice versa, as a probe to establish and symmetry
restrict the ground state magnetic structures of long-range ordered
antiferromagnets.

Regarding the strong correlated system, we would discuss the
spin liquid material as an example. RuCl3 has a rich magnetic phase diagram,
with a complex zig-zag type AFM long-range order at low temperatures,
and even a quantum spin liquid phase in applied magnetic field.
According to our work, non-magnetic and collinearly ordered phases
have vanishing SHE without SOC. For the quantum spin liquid phase,
it remains unclear whether SHE appears without SOC. If yes,
SHE would be a promising probe to the spin liquid phase. We
will study this very interesting question in the future.

\section{method}
To calculate the SHE in these compounds we obtain the DFT Bloch wave functions from Vienna $ab-initio$ Simulation Package (\textsc{vasp})~\cite{kresse1996}
within the generalized gradient approximation (GGA).~\cite{perdew1996}
By projecting the Bloch wave functions onto maximally localized Wannier
functions (MLWFs),~\cite{Mostofi2008} we get a tight-binding Hamiltonian which we use for efficient evaluation of the SHC.
For the integrals of Eq. \ref{SHC}, the BZ was sampled by $k$-grids
from $50\times50\times50$ to $200\times200\times200$. Satisfactory convergence
was achieved for a $k$-grid of size $150\times150\times150$ for all three compounds.
Increasing the grid  size to $200\times200\times200$ only varied the SHC by no more than 5\%.
The results of the calculations agree well with the symmetry analysis.

\section{acknowledgments}
We thank Prof. Claudia Felser at Max Planck Institute in Dresden and Prof. Carsten Timm
at the Technical University Dresden, Prof. Yuval Oreg in the Weizmann Institute of
Science for helpful discussions. Y.Z. and J.v.d.B. acknowledge the German Research
Foundation (DFG) SFB 1143. Y.Z. and J.Z. acknowledge the the ASPIN project (EU FET Open
RIA Grant No. 766566 grant (ASPIN)).
B.Y. acknowledges the support by the Ruth and Herman Albert Scholars Program for New Scientists in Weizmann Institute of Science and by a Grant from the GIF, the German-Israeli Foundation for Scientific Research and Development.
\section{author contributions}
B.Y. conceived the project.
Y.Z., J.Z. and Y.S. performed the DFT calculation and the data analysis.
Y.Z., J.Z., and B.Y. performed the tight-binding modeling and symmetry analysis.
Y.Z. wrote the manuscript with contributions from all co-authors.
J.v.d.B. and B.Y. supervised the work.
We acknowledge the support by ASPIN ( EU FET Open RIA Grant No. 766566).
\section{competing interests}
The authors declare no competing financial interests.


%

\end{document}